\documentstyle[12pt]{article}

\def\a{\alpha }
\def\L{\Lambda }

\begin{document}
\title{{ The
QCD Analysis of the Structure Functions and Effective Nucleon Mass.
}          }

\author{
A.V.Sidorov
\hspace{1mm} \\
Bogoliubov Theoretical Laboratory \\Joint Institute for Nuclear
Research\\
141980 Dubna Russia\\[5mm]
}
\maketitle

\begin{abstract}
{
On the basis of the target mass corrections to structure functions
of deep-inelastic
scattering of leptons, we evaluate effective
nucleon mass that turns out to be twice $M_{nucl.}$
for deep-inelastic
scattering on the nucleus target and equals  $M_{nucl.}$ for the
hydrogen target.
}\\[8mm]
\end{abstract}
\newpage
Deep-inelastic scattering of leptons provides a precise information on
structure functions (SF) of a nucleon.
It is well known that when target mass corrections (TMC) are taken
 into account, the QCD description
 of the SF
 of deep-inelastic scattering is improved. These effect is of the order
$M_{nucl.}^2/Q^2$. In this aricle, we are going to consider the
question wherher the mass of a nucleon is the best value for the
description of data or in order to make the fit better, one has to
use another value $M^{eff.}$  which could differ from the mass of
nucleon.

The Nachtmann moments \cite{Nacht} of SF $F_2$ and $F_3$ are
defined as:\\
\begin{eqnarray}
{}~&&M_2^{QCD}(N,Q^2)=\int_{0}^{1}\frac{dx\xi^{N+1}}{x^3}F_2(x,Q^2)
\frac{3+3(N+1)V+N(N+2)V^2}{(N+2)(N+3)},
\label{f2}
\end{eqnarray}
\begin{eqnarray}
{}~~~~~~M_3^{QCD}(N,Q^2)=\int_{0}^{1}\frac{dx\xi^{N+1}}{x^3}F_3(x,Q^2)
\frac{3+(N+1)V}{(N+2)},
\label{f3}
\end{eqnarray}
where
\begin{eqnarray}
\xi=2x/(1+V),~~~~ V=\sqrt{1+4M_{nucl.}^2x^2/Q^2}
\label{xi}
\end{eqnarray}
Equations (\ref{f2},\ref{f3}) could be expanded into a series in
powers of $M_{nucl.}^2/Q^2$. Retaining only the terms of the order
$M_{nucl.}^2/Q^2$ one could obtain:

\begin{eqnarray}
{}~~~~~M_2(N,Q^2)=M_2^{QCD}(N,Q^2)+\frac{N(N-1)}{N+2}
\frac{M_{nucl.}^2}{Q^2}
M_2^{QCD}(N+2,Q^2),
\label{m2}
\end{eqnarray}

\begin{eqnarray}
{}~~~~~M_3(N,Q^2)=M_3^{QCD}(N,Q^2)+\frac{N(N+1)}{N+2}
\frac{M_{nucl.}^2}{Q^2}
M_3^{QCD}(N+2,Q^2) .
\label{m3}
\end{eqnarray}

$M_2(N,Q^2)$ and $M_2(N,Q^2)$ are the Mellin moments of the
measured SF $F_2$ and $xF_3$:

\begin{eqnarray}
M_2(N,Q^2)=\int_{0}^{1}dx{x^{N-2}}F_{2}(x,Q^2),\\
\label{Mellf2}
M_3(N,Q^2)=\int_{0}^{1}dx{x^{N-2}}xF_{3}(x,Q^2),\\
\label{Mellf3}
 N = 2,3, ...  \nonumber
\end{eqnarray}

  The $Q^2$ - evolution of the moments $ M_2^{QCD}(N,Q^2)$
and $M_3^{QCD}(N,Q^2)$ is  given by QCD \cite{s4,s5}. For the
nonsinglet SF:
\begin{eqnarray}
M_3^{QCD}(N,Q^2)
& =& \left [ \frac{\alpha _{S}\left ( Q_{0}^{2}\right )}
{\alpha _{S}\left ( Q^{2}\right )}\right ]^{d_{N}}
M_3^{QCD}(N,Q^2_0) , \\ \label{m3q2}
&& N = 2,3, ... \nonumber \\
d_N=\gamma^{(0),N}/2\beta_0, && \beta_0=(11-\frac{2}{3}f). \nonumber
\end{eqnarray}

\begin{eqnarray}
\a_s(Q^2)/4\pi&=&1/\beta_0 \ln(Q^2/\L_{LO}^2)
\label{alst}
\end{eqnarray}

\begin{equation}
\gamma^{(0)NS}_{N} ={8\over 3}[1 - {2\over N(N+1)} + 4\sum_{j=2}^{N}
{1\over j}]~.
\label{goa0}
\end{equation}

The unknown coefficients $M_3(N,Q^2_0)$ in (\ref{m3q2}) could be
parametrised as a Mellin moments of some function:

\begin{eqnarray}
M_3^{QCD}(N,Q^2_0)&=&\int_{0}^{1}dx{x^{N-2}}Ax^b(1-x)^c(1+\gamma x),\\
\label{Mellf30}
 && N = 2,3, ...  \nonumber
\end{eqnarray}
where constants A, b, b and $\gamma$ should be determined from the fit
of data.

Having in hand the moments (\ref{Mellf30},\ref{m3q2},\ref{m3},
\ref{Mellf3}) and following the method \cite{Jacobi,Kriv}, we can
write  the structure function $~xF_3~$ in the form:
\begin{equation}
xF_{3}^{N_{max}}(x,Q^2)=x^{\a}(1-x)^{\beta}\sum_{n=0}^{N_{max}}
\Theta_n ^{\a , \beta}
(x)\sum_{j=0}^{n}c_{j}^{(n)}{(\a ,\beta )}
M_{j+2}^{NS} \left ( Q^{2}\right ),   \\
\label{e7}
\end{equation}
where $~\Theta^{\alpha \beta}_{n}(x)~$ is a set of Jacobi polynomials
and $~c^{n}_{j}(\alpha,\beta)~$ are coefficients of the series of
$~\Theta^{\alpha,\beta}_{n}(x)~$ in powers of x:
\begin{equation}
\Theta_{n} ^{\a , \beta}(x)=
\sum_{j=0}^{n}c_{j}^{(n)}{(\a ,\beta )}x^j .
\label{e9}
\end{equation}

The quantities $N_{max},~ \alpha~$ and $~\beta~$ have to be chosen
so as to achieve the most fast convergence of the series in the r.h.s.
of Eq.(\ref{e7}) and to
reconstruct $~xF_3~$ with
the accuracy required. Following the results of \cite{Kriv}
we use $~\alpha = 0.12~,
{}~\beta = 2.0~$ and $~N_{max} = 12~$. These numbers guarantee accuracy
better than $~10^{-3}~$.\\

Eq. (\ref{e7}) could be applied for reconstructing  SF $F_2(x,Q^2)$ for
$0.3\leq x$ and with eq. (\ref{f2},\ref{m2}) for TMC taken into account.

The parameters A, b, c, $\gamma$ and parameter $\Lambda$ are determined
by fitting
experimental data. We also consider $M^{eff.}$ as a free parameter.
It should be noted that the parameters a, b, c and  $\gamma$  depend
on $Q^2_0$. We have used experimental points with $Q^2>5 GeV^2$ for
fitting, in order to avoid high--twist effects and chose $Q^2_0=10
GeV^2$.

The results of concrete calculations made for SF measured in
experiments on different targets are presented in Table I.\\

For the hydrogen target $M^{eff.}$ reproduces the value of the proton
mass. For the iron target the effective mass  $M^{eff.}$ is twice
the nucleon mass.
The data of the SKAT collaboration on a target which consists
of a mixture of Neon and
Hydrogen are not precise enough to determine the value of $\Lambda$.
So following \cite{SKAT} we have fixed $\Lambda=200 MeV$ and found
the value of $M^{eff.}$ a little bit higher than for the hydrogen
target. The increasing effective mass of a nuclon
on the nucleus target takes place for a
nonsinglet fit both for $F_2$ and $xF_3$ SF. It also takes place
both for the leading and next to leading order QCD
(see result for $xF_3$ data of CCFR Table 1.).
The large value of $M^{eff.}$ found in the QCD fit of data
of DIS on nucleon
target could be considered as indirect evidence of the existence of
multiquark clusters   \cite{Bal,Krz,Kon,Fuj} or a few--nucleon
correlation in a nucleus \cite{Fra}.  It is also compatible with the
measured SF at $x>1$  on DIS of leptons on the nucleus target
\cite{BCDMSx1}.\\


We are thankful to Profs. A.E. Dorokhov, S.B. Gerasimov,
A.L.~Kataev, N. Stefanis and M.V. Tokarev for fruitful discussions.\\

This research was partly supported by INTAS (International
Association for the Promotion of Cooperation with Scientists from the
Independent States of the Former Soviet Union) under Contract
nb 93-1180, by the Heisenberg--Landau Program and by
the Russian Fond for Fundamental Research Grant N 94-02-04548-a.

\newpage
\begin{tabular}{|lll|c|c|c|c|c|} \hline
\multicolumn{3}{|c|}{ Collaboration}  &  Ref.    &            &
 $\Lambda$
                                       & $\chi^2_{d.f.}$ & $M^{eff.}$\\
\multicolumn{3}{|c|}{ Reaction }      &               &
&$[MeV]$&  &  $[GeV]$    \\  \hline
BCDMS~ $\mu p$ ~~  $ F_2        $&&&\cite{BCDMS}  &$ 0.35 < x$  &130
$ \pm$ 4  & 183/223 & 0.88 $ \pm$ 0.14      \\
SKAT  $\nu Ne,p$~~$ xF_3       $&&&\cite{SKAT} &$ 0.05 \leq x$  &200
(fix.)   & 25.3/30 & 1.42 $\pm$ 0.71     \\
EMC   $\mu Fe$~~    $ F_2        $&&&\cite{EMC}    &$ 0.30 < x $ &106
 $ \pm$ 26 &45.3/45  &  2.08 $ \pm$ 0.16     \\
CCFR  $\nu Fe$~~    $ F_2~~    $&&& \cite{f3data} &$ 0.275 \leq x$ &
146 $\pm$ 12   & 37.9/81 & 1.76 $\pm$ 0.09 \\
CCFR  $\nu Fe$~~    $ xF_3~~   $&&& \cite{f3data} &$ 0.015 \leq x$ &
64.7 $\pm$ 21   & 81.8/81 & 2.04 $\pm$ 0.18  \\
CCFR  $\nu Fe$~~    $xF_3~~  NLO $&&& \cite{f3data} &$ 0.015 \leq x$
& 116 $\pm$ 30   & 73.4/81 & 1.83 $\pm$ 0.20   \\
\hline
\multicolumn{8}{l}{{\bf Table I.} The summary of various
determinations of the $M_n^{eff.}$.
}
\\[5mm]
\end{tabular}


\newpage
\vskip 1cm
\begin{thebibliography}{99}
\bibitem{Nacht}  O. Nachtmann, Nucl. Phys. {\bf B63} (1973) 237; \\
S. Wandzura, Nucl. Phys. {\bf B122} (1977) 412.
\vspace*{-2mm}
\bibitem{s4} F.J.Yndurain, Quantum Chromodynamics (An Introduction  to
the Theory of Quarks and Gluons).- Berlin, Springer-Verlag
(1983), 117.
 \vspace*{-2mm}
\bibitem{s5}  A.Buras, Rev. Mod. Phys. {\bf 52} (1980) 199.
 \vspace*{-2mm}
\bibitem{Jacobi}
G. Parisi, N. Sourlas,
\newblock {   Nucl. Phys.} {\bf B151} (1979) 421;
\\ I. S. Barker, C. B. Langensiepen, G. Shaw,
\newblock {   Nucl. Phys.} {\bf B186} (1981) 61;\\
I.S.Barker, B.R.Martin, G.Shaw, Z. Phys. {\bf C19} (1983)  147; \\
I.S.Barker, B.R.Martin, Z. Phys. {\bf C24} (1984) 255; \\
S.P.Kurlovich, A.V.Sidorov, N.B.Skachkov, JINR Report {\bf E2-89-655},
Dubna, 1989.
\vspace*{-2mm}
\bibitem{Kriv}
V. G. Krivokhizhin et al.,
\newblock {   Z. Phys.} {\bf C36} (1987) 51;
\\V. G. Krivokhizhin et al.,
\newblock {   Z. Phys.} {\bf C48} (1990) 347;\\
A. L. Kataev and A.V. Sidorov,
\newblock {   Phys. Lett.} {\bf B331} (1994) 179.
\bibitem{BCDMS}
BCDMS Collab., A. Benvenuti et al.,
\newblock{   Phys. Lett.} {\bf 223B} (1989) 485.
\vspace*{-2mm}
\bibitem{SKAT} SKAT Coll., V.V. Ammosov et al., Z. Phys. {\bf C30}
(1986) 175.
\vspace*{-2mm}
\bibitem{EMC} J.J.Aubert  et al., Phys. Lett., {\bf 105B} (1982) 322.
 \vspace*{-2mm}
\bibitem{f3data}
CCFR Collab., S. R. Mishra et al.,
\newblock Nevis Preprint {\bf N 1459} (1992);
CCFR Collab., W. C. Leung et al.,
\newblock {   Phys. Lett.} {\bf B317} (1993) 655;
CCFR Collab., P. Z. Quintas et al.,
\newblock {   Phys. Rev. Lett.} {\bf 71} (1993) 1307.
 \vspace*{-2mm}
\bibitem{Bal} A. M. Baldin et al., Proc. Rochester Meeting APS/DPF,
New York 1971, pp 131-137;\\
A. M. Baldin et al., Sov. J. Nucl. Phys. {\bf 18} (19740 41.
 \vspace*{-2mm}
\bibitem{Krz}  A. Krzywicki, Nucl. Phys. {\bf A446} (1985) 135.
 \vspace*{-2mm}
\bibitem{Kon} L. A. Kondratyuk and M. Zh. Shmatikov, Yad. Fiz. {\bf 41}
 (1985) 222;
Letters to ZhETF {\bf 39} (1984) 324.
 \vspace*{-2mm}
\bibitem{Fuj} T. Fujita, J. Hufner, Nucl. Phys. {\bf A343} (1980) 493.
 \vspace*{-2mm}
\bibitem{Fra} L. Frankfurt and M. Strikman, Phys. Rep. {\bf 76} (1981)
216.
 \vspace*{-2mm}
\bibitem{BCDMSx1}
BCDMS Collab., A. Benevenuti et al.,
\newblock {\it Z. Phys.} {\bf C63} (1993) 29.
\vspace*{-2mm}
\end{thebibliography}
\end{document}